\documentclass[graybox]{svmult}

\usepackage{mathptmx}       
\usepackage{helvet}         
\usepackage{courier}        
\usepackage{type1cm}        
\usepackage{makeidx}         
\usepackage{graphicx}        
\usepackage{multicol}        
\usepackage[bottom]{footmisc}
\usepackage{amssymb}

\makeindex             

\begin{document}

\title*{A double complex construction and discrete Bogomolny equations }
\author{Volodymyr Sushch}
\institute{Volodymyr Sushch \at Koszalin University of Technology, Sniadeckich 2,
 75-453 Koszalin, Poland, \email{volodymyr.sushch@tu.koszalin.pl}}

%
%
\maketitle

\abstract*{We study discrete models  which are generated by the  self-dual Yang-Mills equations.
 Using a double complex construction we construct a new discrete analog of the
Bogomolny equations. Discrete Bogomolny equations, a system of matrix valued
difference equations, are obtained from discrete self-dual equations.}

\abstract{We study discrete models which are generated by the  self-dual Yang-Mills equations. Using a double complex construction we construct a new discrete analog of the Bogomolny equations. Discrete Bogomolny equations, a system of matrix valued
difference equations, are obtained from discrete self-dual equations.}

\section{Introduction}
\label{sec:1}
This  work  is concerned  with
discrete model of the $SU(2)$ self-dual Yang-Mills equations
described in \cite{S2}. It is well known that the self-dual Yang-Mills equations admit reduction to the
Bogomolny equations \cite{AH}. Let $A$ be an $SU(2)$-connection on   $\mathbb{R}^3$. This means that $A$ is an
$su(2)$-valued 1-form and we can write
\begin{equation}\label{eq:01}
A=\sum_{i=1}^3A_i(x) \D x^i,
\end{equation}
where \ $A_i: \mathbb{R}^3\rightarrow su(2)$.  Here $su(2)$ is the Lie algebra of $SU(2)$. The connection $A$  is  also called a gauge
potential with the gauge group $SU(2)$ (see \cite{NS} for more details). Given  the connection $A$ we define the curvature 2-form $F$ by
\begin{equation} \label{eq:02}
 F=\D A+A\wedge A,
 \end{equation}
 where $\wedge$ denotes the exterior multiplication of differential forms. Let $\varPhi: \mathbb{R}^3\rightarrow su(2)$ be a scalar field (a Higgs field).
The Bogomolny equations are a set of nonlinear
partial differential
 equations, where unknown is a pair ($A, \varPhi$).
  These equations can be written as
\begin{equation}\label{eq:03}
F=\ast \D_A \varPhi,
\end{equation}
where   $\ast$ is the Hodge star operator on
$\mathbb{R}^3$ and \ $\D_A$ is the covariant exterior differential
operator. This operator is defined by the formula
\begin{eqnarray*}
\D_A\varOmega=\D\varOmega+A\wedge \varOmega+(-1)^{r+1}\varOmega\wedge A,
\end{eqnarray*}
 where  $\varOmega$ is an arbitrary   $su(2)$-valued  $r$-form.

 Let now consider the connection $A$ on $\mathbb{R}^4$. We define $A$ to be
\begin{equation}\label{eq:04}
A=\sum_{i=1}^3A_i(x) \D x^i+\varPhi(x)\D x^4,
\end{equation}
where  $A_i$ and $\varPhi$ are independent of $x^4$. In other word, the scalar field  $\varPhi$ is identified with a fourth component $A_4$ of the connection $A$. It is easy to check that if the pair ($A, \varPhi$) satisfies  Equation~(\ref{eq:03}), then the connection (\ref{eq:04}) is a solution of the self-dual equation
\begin{equation}\label{eq:05}
 F=\ast F.
 \end{equation}
  In fact, the Bogomolny equations can be obtained  from the self-dual equations by using dimensional reduction from $\mathbb{R}^4$ to $\mathbb{R}^3$ \cite{AH}.

  The aim of this paper  is  to construct a
 discrete model of Equation~(\ref{eq:03}) that preserves the geometric structure of
the original continual object. This mean that
speaking of a discrete model, we  mean not only  the direct
replacement of differential operators by  difference ones  but
also  a discrete analog of the Riemannian structure over a
properly introduced  combinatorial object. The idea presented here is strongly influenced by book Dezin \cite{Dezin}. Using a double complex construction we construct a new discrete analog of the Bogomolny equations. In much the same way as in the continual case these discrete equations are obtained from discrete self-dual equations. We continue the investigations \cite{S1, S2}, where discrete analogs of the self-dual and anti-self-dual equations on a double complex are studied. It should be noted that there are many other approaches to discretisation of Yang-Mills theories. As the list of papers on the subject is very large, we content ourselves by referencing the works \cite{CC, GN, KSSB, KT, MS, O}. In these papers some other discrete versions of the Bogomolny equations are studied.

\section{Double complex construction}
\label{sec:2}
The double complex construction is described in \cite{S1}. For the convenience of the reader we briefly repeat the relevant material from  \cite{S1}
without proofs.
Let the tensor product \  $C(n)=C\otimes ... \otimes C$
of an  1-dimensional complex $C$ be a combinatorial model of
Euclidean space $\mathbb{R}^n$.
The 1-dimensional complex $C$ is defined in the following way. Let
\ $C^0$ denotes the real linear space of 0-dimensional chains
generated by basis elements \ $x_i$ (points), $i\in
\mathbb{Z}$. It is convenient to introduce the shift operator
$\utau$ in the set of indices by
\begin{eqnarray*}
\utau i=i+1.
 \end{eqnarray*}
We denote the open interval
$(x_i, \ x_{\utau i})$ by \ $e_i$. We can regard the
set $\{e_i\}$ as a set of basis elements of the real linear
space \ $C^1$  of 1-dimensional chains. Then the 1-dimensional
complex (combinatorial real line) is the direct sum of the
introduced spaces \ $C=C^0\oplus C^1$. The boundary operator
\ $\partial$ \ on the basis elements of \ $C$ is given by
 \begin{equation}\label{eq:06}
 \partial x_i=0, \qquad  \partial
e_i=x_{\utau i}-x_i.
 \end{equation}
 The definition is extended to arbitrary chains by linearity.

 Multiplying the basis elements \ $x_i$ and \ $e_i$ \ of \  $C$ in various way
we obtain basis elements of \ $C(n)$.
Let \  $s_k^{(r)}=s_{k_1}\otimes ... \otimes s_{k_n}$, where \ $k=(k_1,...,k_n)$ and \ $k_i\in\mathbb{Z},$ \ be an arbitrary
$r$-dimensional basis element of $C(n)$. The product
contains exactly $r$ of $1$-dimensional elements \ $e_{k_i}$ and \ $n-r$ \ of
 $0$-dimensional elements \  $x_{k_i}$. The superscript $(r)$ also uniquely determines an $r$-dimensional basis element of $C(n)$.
For example, the 1-dimensional \  $e_k^i$ and 2-dimensional \ $\varepsilon_k^{ij}$ basis elements
of $C(3)$ can be written as
\begin{eqnarray*}
 e_k^1=&e_{k_1}\otimes
x_{k_2}\otimes x_{k_3}, \quad e_k^2=
x_{k_1}\otimes e_{k_2}\otimes x_{k_3}, \quad e_k^3&=
x_{k_1}\otimes x_{k_2}\otimes e_{k_3},\\
\varepsilon_k^{12}=&e_{k_1}\otimes
e_{k_2}\otimes x_{k_3}, \quad
\varepsilon_k^{13}=e_{k_1}\otimes x_{k_2}\otimes e_{k_3}, \quad \varepsilon_k^{23}&=x_{k_1}\otimes e_{k_2}\otimes e_{k_3},
 \end{eqnarray*}
where \ $k=(k_1,k_2,k_3)$ and \ $k_i\in\mathbb{Z}.$

Now we  consider a dual object of the  complex  $C(n)$. Let $K(n)$
be a cochain complex with  $gl(2,\mathbb{C})$-valued coefficients,
where  $gl(2,\mathbb{C})$ is the Lie algebra of the group
$GL(2,\mathbb{C})$. We suppose that the complex $K(n)$, which is a
conjugate of $C(n)$, has a similar structure: ${K(n)=K\otimes
... \otimes K}$, where $K$ is a dual of the 1-dimensional
complex $C$. We will write the basis elements of $K$ as ${x^i},
\ {e^i}$. Then an arbitrary basis element of $K(n)$ is given by
${s^k= s^{k_1}\otimes ... \otimes s^{k_n}}$, where
$s^{k_i}$ is either $x^{k_i}$ or $e^{k_i}$.
For an $r$-dimensional cochain $\varphi\in K(n)$ we have
\begin{equation}\label{eq:07}
\varphi= \sum_k \sum_r\varphi_k^{(r)}s_{(r)}^k,
 \end{equation}
where $\varphi_k^{(r)}\in gl(2,\mathbb{C})$.
We will
call cochains forms, emphasizing their relationship with the
corresponding continual objects, differential forms.

We define the pairing operation for arbitrary
basis elements \ $\varepsilon_k\in C(n)$,  $s^k\in K(n)$ by the
rule
\begin{equation}\label{eq:08}
 <\varepsilon_k, \ as^k>=\left\{\begin{array}{l}0,\quad
\varepsilon_k\ne s_k\\
                            a, \quad \varepsilon_k=s_k, \quad a\in gl(2,\mathbb{C}).
                            \end{array}\right.
\end{equation}
Here for simplicity  the superscript $(r)$ is omitted.
 The operation (\ref{eq:08}) is linearly extended to cochains.

The operation $\partial$  induces the dual operation
\ $\D ^c$ on $K(n)$ in the following way:
\begin{equation}\label{eq:09}
<\partial\varepsilon_k, \ as^k>=<\varepsilon_k, \ a\D^cs^k>.
\end{equation}
For example, if $\varphi$ is a 0-form, i.e.
$\varphi= \sum_k \varphi_kx^k,$
 where $x^k=x^{k_1}\otimes ... \otimes x^{k_n}$, then
\begin{equation}\label{eq:10}
\D^c\varphi= \sum_k \sum_{i=1}^n(\varDelta_i\varphi_k)e_i^k,
\end{equation}
where  \ $e_i^k$ \ is the 1-dimensional basis elements of $K(n)$ and
\begin{equation}\label{eq:11}
\varDelta_i\varphi_k=\varphi_{\utau_ik}-\varphi_k.
\end{equation}
Here the shift operator \ $\utau_i$ acts as
 \begin{eqnarray*}
 \tau_ik=(k_1, ..., \utau k_i, ..., k_n).
 \end{eqnarray*}
 The coboundary operator $\D^c$ is an analog of the  exterior
differentiation operator \  $\D$.

Introduce a cochain product on  $K(n)$.
  We denote this product by
 $\cup$. In terms of the homology theory this is the so-called Whitney product.
  For the basis elements of 1-dimensional complex $K$ the
$\cup$-product is defined as follows
\begin{eqnarray*}
 x^i\cup
x^i=x^i, \quad e^i\cup x^{\utau i}=e^i,
\quad x^i\cup e^i=e^i, \quad i\in\mathbb{Z},
\end{eqnarray*}
supposing the product to be zero in all other case.
By induction we extend this definition to basis elements of $K(n)$ (see \cite{S1} for details).
For example, for the 1-dimensional   basis elements
 \ $e^k_i\in K(3)$ we have
\begin{eqnarray}\nonumber
e_1^k\cup e_2^{\utau_1k}=&\varepsilon^k_{12},
\qquad e_1^k\cup e_3^{\utau_1k}=\varepsilon^k_{13}, \qquad e_2^k\cup e_3^{\utau_2k}&=\varepsilon^k_{23},\\ \label{eq:12}
e_2^k\cup e_1^{\utau_2k}=&-\varepsilon^k_{12}, \ \quad e_3^k\cup e_1^{\utau_3k}=-\varepsilon^k_{13}, \ \quad e_3^k\cup e_2^{\utau_3k}&=-\varepsilon^k_{23}.
\end{eqnarray}
To arbitrary
forms the $\cup$-product be extended linearly.
Note that the coefficients of forms
multiply as matrices.
It is worth pointing out that for any  forms  \ $\varphi,  \psi\in K(n)$ the following relation holds
\begin{equation}\label{eq:13}
 \D^c(\varphi\cup\psi)=\D^c\varphi\cup\psi+(-1)^r\varphi\cup\D^c\psi,
\end{equation}
 where  $r$  is the dimension of a form
$\varphi$. For the proof we refer the reader to \cite{Dezin}.
Relation (\ref{eq:13}) is a discrete analog of the Leibniz rule for differential forms.

Let us now together with the complex $C(n)$  consider its "double", namely the
complex $\tilde{C}(n)$ of exactly the same structure. Define the
one-to-one correspondence
\begin{equation}\label{eq:14}
\ast : C(n)\rightarrow\tilde{C}(n), \qquad \ast : \tilde
C(n)\rightarrow C(n)
\end{equation}
in the following way:
\begin{equation}\label{eq:15}
\ast : s_k^{(r)}\rightarrow\pm\tilde s_k^{(n-r)}, \qquad \ast :
\tilde s_k^{(r)}\rightarrow \pm s_k^{(n-r)},
\end{equation}
where
 \ $\tilde s_k^{(n-r)}=*s_{k_1}\otimes ... \otimes *s_{k_n}$
and \ $*s_{k_i}=\tilde e_{k_i}$ if  $s_{k_i}=x_{k_i}$ and
 $*s_{k_i}=\tilde x_{k_i}$ if  $s_{k_i}=e_{k_i}.$
We let the plus sign  in (\ref{eq:15}) if a permutation of $(1, ..., n)$ with $(1, ..., n)\rightarrow ((r), ...,(n-r))$ is representable as the product
of an even number of transpositions and  the minus sign  otherwise.

The complex of the cochains $\tilde K(n)$ over the double complex
$\tilde C(n)$ has the same structure as  $K(n)$. Note that  forms $\varphi\in K(n)$ and   $\tilde\varphi\in\tilde K(n)$ have both the same components.
The operation
(\ref{eq:14}) induces the respective mapping
\begin{equation}\label{eq:16}
\ast : K(n)\rightarrow\tilde{K}(n), \qquad \ast : \tilde
K(n)\rightarrow K(n)
\end{equation}
by the rule:
\begin{eqnarray*}
<\tilde
c, \ *\varphi>=<*\tilde c, \ \varphi>, \qquad <c, \ *\tilde
\psi>=<*c, \ \tilde\psi>,
\end{eqnarray*}
where \ $c\in C(n), \ \tilde c\in\tilde C(n), \ \varphi\in K(n), \
\tilde\psi\in\tilde K(n)$.
For example, for the 2-dimensional   basis elements \
 $\varepsilon^k_{ij}\in K(3)$ we have
 \begin{equation}\label{eq:17}
 \ast\varepsilon^k_{12}=\tilde e^k_3, \quad \ast\varepsilon^k_{13}=-\tilde e^k_2, \quad \ast\varepsilon^k_{23}=\tilde e^k_1.
 \end{equation}
 This operation is a discrete analog of the Hodge star operation.
Similarly to the continual case we have
\begin{eqnarray*}
\nonumber\ast\ast\varphi=(-1)^{r(n-r)}\varphi
\end{eqnarray*}
 for any discrete $r$-form
$\varphi\in K(n)$.

Finally, for convenience we introduce the following operation
\begin{equation}\label{eq:18}
\tilde\iota: K(n)
\rightarrow \tilde K(n), \qquad \tilde\iota: \tilde K(n)
\rightarrow K(n)
\end{equation}
by setting
 $ \ \tilde\iota s_{(r)}^k= \tilde s_{(r)}^k, \quad \tilde\iota\tilde s_{(r)}^k=  s_{(r)}^k.$
It is easy to check that  the following hold
\begin{eqnarray*}
 \tilde\iota\ast&=\ast\tilde\iota, \quad
\tilde\iota d^c=d^c\tilde\iota, \quad \tilde\iota\varphi=\tilde\varphi, \quad \tilde\iota\tilde\iota\varphi=\varphi,  \quad
\tilde\iota(\varphi\cup\psi)&=\tilde\iota\varphi\cup\tilde\iota\psi,
\end{eqnarray*}
where \ $\varphi,  \psi\in K(n)$.

\section{Discrete Bogomolny equations}
\label{sec:3}
Let us consider a discrete 0-form $\varPhi\in K(3)$ with coefficients belonging to
$su(2)$. We put
\begin{equation}\label{eq:19}
\varPhi=\sum_k \varPhi_kx^k,
\end{equation}
where $\varPhi_k\in su(2)$  and  $x^k=x^{k_1}\otimes x^{k_2}\otimes x^{k_3}$ is the
 0-dimensional basis element of  $K(3)$, \  $k=(k_1,k_2,k_3), \
k_i\in\mathbb{Z}.$

 We define a discrete $SU(2)$-connection  $A$ to be
\begin{equation}\label{eq:20}
A=\sum_k\sum_{i=1}^3A_k^ie_i^k,
\end{equation}
where  $A_k^i\in su(2)$  and  \ $e_i^k$ is the
 1-dimensional basis element of  $K(3)$.

On account of (\ref{eq:07}) an arbitrary discrete 2-form \ $F\in K(3)$  can be written  as follows
\begin{equation}\label{eq:21}
 F=\sum_k\sum_{i<j} F_k^{ij}\varepsilon_{ij}^k=\sum_k\big(F_k^{12}\varepsilon_{12}^k+F_k^{13}\varepsilon_{13}^k+F_k^{23}\varepsilon_{23}^k\big),
 \end{equation}
 where $ F_k^{ij}\in gl(2,\mathbb{C})$  and   $ \varepsilon_{ij}^k$  is the
  2-dimensional  basis element of  $K(3)$.
Define a  discrete analog of the curvature form  (\ref{eq:02}) by
\begin{equation}\label{eq:22}
F=\D^cA+A\cup A.
\end{equation}
By the definition of $\D^c$ (\ref{eq:09}) and using (\ref{eq:12}) we have
\begin{equation}\label{eq:23}
 \D^cA=\sum_k\sum_{i<j}(\varDelta_iA_k^j-\varDelta_jA_k^i)\varepsilon_{ij}^k
 \end{equation}
 and
 \begin{equation}\label{eq:24}
 A\cup A=\sum_k\sum_{i<j}(A_k^iA_{\utau_ik}^j-
 A_k^jA_{\utau_jk}^i)\varepsilon_{ij}^k.
 \end{equation}
 Recall that  $\varDelta_i$ is the difference operator (\ref{eq:11}).
  Combining (\ref{eq:23}) and (\ref{eq:24}) with (\ref{eq:21}) we obtain
\begin{equation}\label{eq:25}
 F_k^{ij}=\varDelta_iA_k^j-\varDelta_jA_k^i+A_k^iA_{\utau_ik}^j-
 A_k^jA_{\utau_jk}^i.
 \end{equation}

 It should be noted that in the continual case the curvature form $F$
  takes values in the algebra \ $su(2)$ for any $su(2)$-valued connection form $A$.
 Unfortunately, this is not true in the discrete case because, generally speaking, the components
  \ $A_k^iA_{\utau_ik}^j- A_k^jA_{\utau_jk}^i$ of the form $A\cup A$ in (\ref{eq:22}) do not belong to $su(2)$.
 For a definition of the $su(2)$-valued discrete curvature form we refer the reader to \cite{S2}.

 Let us define a discrete analog of the exterior covariant differential operator $\D_A$ as follows
\begin{eqnarray*}
\D_A^c\varphi=\D^c\varphi+A\cup\varphi+(-1)^{r+1}\varphi\cup A,
\end{eqnarray*}
where $\varphi$ is an arbitrary $r$-form (\ref{eq:07}) and $A$ is given by (\ref{eq:20}) .
Then for the 0-form (\ref{eq:19}) we obtain
\begin{equation}\label{eq:26}
\D^c_A\varPhi=\D^c\varPhi+A\cup\varPhi-\varPhi\cup A.
\end{equation}
Using (\ref{eq:10}) and the definition of $\cup$ we can rewritten (\ref{eq:26}) as follows
\begin{equation}\label{eq:27}
\D^c_A\varPhi=\sum_k\sum_{i=1}^3(\varDelta_i\varPhi_k+A_k^i\varPhi_{\utau_ik}-\varPhi_kA_k^i)e_i^k.
\end{equation}
Applying the operation $\ast$ (\ref{eq:16}) to this expression and by (\ref{eq:17})  we find
\begin{eqnarray}\nonumber
\ast\D^c_A\varPhi=\sum_k(\varDelta_1\varPhi_k+A_k^1\varPhi_{\utau_1k}-\varPhi_kA_k^1)\tilde\varepsilon_{23}^k\\ \nonumber
-\sum_k(\varDelta_2\varPhi_k+A_k^2\varPhi_{\utau_2k}-\varPhi_kA_k^2)\tilde\varepsilon_{13}^k \\
+\sum_k(\varDelta_3\varPhi_k+A_k^3\varPhi_{\utau_3k}-\varPhi_kA_k^3)\tilde\varepsilon_{12}^k.\label{eq:28}
\end{eqnarray}
Now suppose that $\varPhi$ in the form (\ref{eq:19}) is a discrete analog of the  Higgs field. Then the discrete analog of the Bogomolny equation (\ref{eq:03})  is given by the formula
\begin{equation}\label{eq:29}
 F=\tilde\iota\ast \D_A^c\varPhi,
\end{equation}
where \ $\tilde\iota$ \ is the operation (\ref{eq:17}). From (\ref{eq:21}) and (\ref{eq:28}) it follows immediately that
Equation (\ref{eq:29}) is equivalent to the following difference equations
\begin{eqnarray}\nonumber
 F_k^{12}=&\varDelta_3\varPhi_k+A_k^3\varPhi_{\utau_3k}-
 \varPhi_kA_k^3, \\ \nonumber
  F_k^{13}=&-\varDelta_2\varPhi_k-A_k^2\varPhi_{\utau_2k}+
 \varPhi_kA_k^2, \\
 F_k^{23}=&\varDelta_1\varPhi_k+A_k^1\varPhi_{\utau_1k}-
 \varPhi_kA_k^1.\label{eq:30}
 \end{eqnarray}

 Consider now the discrete curvature form  (\ref{eq:22}) in the  4-dimensional case, i. e. $F\in K(4)$.
 The discrete analog of the  self-dual  equation  (\ref{eq:05}) can be written as
follows
\begin{equation}\label{eq:31}
 F=\tilde\iota\ast F.
\end{equation}
By the definition of  $\ast$ for the 2-dimensional basis elements $\varepsilon^k_{ij}\in K(4)$ we have
\begin{eqnarray*}
 \ast \varepsilon^k_{12}=\tilde\varepsilon^k_{34}, \quad  \ast\varepsilon^k_{13}=-\tilde\varepsilon^k_{24}, \quad  \ast\varepsilon^k_{14}=\tilde \varepsilon^k_{23}, \\
 \ast \varepsilon^k_{23}=\tilde\varepsilon^k_{14}, \quad  \ast\varepsilon^k_{24}=-\tilde\varepsilon^k_{13}, \quad  \ast\varepsilon^k_{34}=\tilde \varepsilon^k_{12}.
 \end{eqnarray*}
 Using this we may compute $\ast F$:
 \begin{eqnarray*}
 \ast F=\sum_k\big(F_k^{12}\tilde\varepsilon^k_{34}-F_k^{13}\tilde\varepsilon^k_{24}+F_k^{14}\tilde \varepsilon^k_{23}+F_k^{23}\tilde\varepsilon^k_{14}-
F_k^{24}\tilde\varepsilon^k_{13}+F_k^{34}\tilde \varepsilon^k_{12}\big).
 \end{eqnarray*}
 Then  Equation (\ref{eq:31}) becomes
\begin{equation}\label{eq:32}
 F_k^{12}=F_k^{34}, \qquad F_k^{13}=-F_k^{24}, \qquad
 F_k^{14}=F_k^{23}.
\end{equation}
Let the discrete connection 1-form $A\in K(4)$ is given by
\begin{equation}\label{eq:33}
A=\sum_k\sum_{i=1}^3A_k^ie_i^k+\sum_k\varPhi_ke_4^k,
\end{equation}
where  $A_k^i\in su(2),$ \ $\varPhi_k\in su(2)$  and  \ $k=(k_1,k_2,k_3,k_4),$ \ $k_i\in\mathbb{Z}.$
Note that  here we put \ $A_k^4=\varPhi_k$ and $\varPhi_k$ are the components of the discrete Higgs field.
Suppose that the connection form (\ref{eq:33}) is independent of \ $k_4$, i.e.
 \begin{equation}\label{eq:34}
\varDelta_4A_k^i=0, \qquad   \varDelta_4\varPhi_k=0
\end{equation}
for any \ $i=1,2,3$ \ and \ $k=(k_1,k_2,k_3,k_4)$.
Substituting (\ref{eq:34}) into (\ref{eq:25}) yields
\begin{eqnarray*}
 F_k^{i4}=\varDelta_i\varPhi_k+A_k^i\varPhi_{\utau_ik}-
 \varPhi_kA_k^i,  \qquad i=1,2,3.
\end{eqnarray*}
 Putting these expressions  in Equations (\ref{eq:32}) we obtain  Equations (\ref{eq:30}).

 Thus, if  the component $A_k^4$ of $A$
 is identified with $\varPhi_k$ for any $k=(k_1,k_2,k_3,k_4),$ \ $k_i\in\mathbb{Z}$, then the discrete Bogomolny equations and the discrete  self-dual  equations are equivalent.

\end{document}